\documentclass[%
preprint,
amsmath,amssymb,
aps
floatfix,
]{revtex4-2}

\usepackage{graphicx}
\usepackage{dcolumn}
\usepackage{bm}
\usepackage{float}

\begin{document}

	\title{Impact of a CSS quantum error correction code in underwater quantum key distribution}

	\author{Juliette FLORIN}
	\affiliation{Thales, Brest, France }
	\affiliation{
		Institut Polytechnique de Paris
	}%
	\affiliation{
		EURECOM, 450 Route des Chappes, 06410 Biot, France
	}%
	\email{juliette.florin@eurecom.fr}
	
	\author{Nicolas LE JOSSE}%
	\affiliation{%
		Thales, Brest, France 
	}%
	\author{Arnaud COATANHAY }
	\affiliation{
		Lab-STICC, UMR CNRS 6285, ENSTA, Institut Polytechnique de Paris, 2 rue François Verny, 29806 Brest Cedex 9, France
	}%
	
	\author{Gilles BUREL}
	\affiliation{%
		Univ. Brest, CNRS, Lab-STICC UMR 6285, 6 av. Le Gorgeu, 29200 Brest, France
	}%

	\date{\today}%

	\begin{abstract}
		Quantum key distribution (QKD) enables secure underwater communications essential for maritime infrastructure. Underwater optical channels introduce substantial photon loss (erasures) and ambient noise that degrade QKD performance. This paper investigates whether two four-qubit Calderbank-Shor-Steane (CSS) quantum error correction codes (QECC) mitigate these impairments in vertical underwater communication BB84 QKD protocol. After developing a comprehensive stochastic channel model incorporating photon loss, geometric spreading, and solar noise, we assess the viability of QECC through the analytical study of the quantum bit error rate (QBER) and the secure key rate (SKR) with and without security depending on the signal-to-noise ratio (SNR) validated against Monte Carlo simulations. The standard four-qubit CSS code achieves a 3 dB SNR gain at the QBER 11\% security threshold; the discard code variant achieves 4.5 dB. However, QECC includes an encoding overhead that reduces the SKR. We demonstrate a crucial relationship between the SKR and the probability of arrival of the sent photon. Analysis shows that QECC is beneficial exclusively in marginal SNR regimes; at high SNR, raw BB84 dominates. For an ocean Type III Jerlov water scenario with sunlight coming from the sun located at the top of the atmosphere, we identify operational depth-range windows where QECC enables communication otherwise infeasible. This analysis establishes that error correction deployment must be scenario-dependent: extend operational range at the cost of throughput when SNR is marginal, or prioritize key generation rates at high SNR.
	\end{abstract}
	
	\maketitle
	\section{\label{sec:Introduction}Introduction}
	
	Secure underwater communications are essential for mission-critical maritime activities \cite{paglierani_primer_2023} like coastline protection, remote control of submarine oil extraction, and marine life monitoring, yet they face unique constraints from the underwater channel  \cite{lal_toward_2017,han_secure_2015}. There is an urgent need to secure critical maritime infrastructure against the escalating threat posed by quantum computing \cite{meena_continuous_2025}. Furthermore, as we move toward a quantum internet \cite{kimble_quantum_2008,pirandola_fundamental_2017}, where nodes linked by quantum channels enable tasks such as the distribution of quantum states, underwater networks remain a vital yet vulnerable frontier. By studying underwater quantum key distribution (QKD), we address this challenge by enabling key agreements, where any eavesdropping attempt unavoidably disturbs quantum signals and becomes detectable \cite{paglierani_primer_2023}.

	QKD in underwater channels presents unique challenges due to high photon loss arising from water absorption and scattering, making it essential to have an accurate channel model in order to identify the dominant error mechanisms and enable appropriate countermeasures \cite{montiel_ross_performance_2025}. Experimental demonstrations of underwater protocol BB84 QKD via polarization encoded photons \cite{feng_experimental_2021, zhao_experimental_2019} have motivated a few theoretical investigations; however, rigorous channel modeling in the literature remains limited \cite{rizk_bbm92_2026,meena_continuous_2025,raouf_performance_2022,paglierani_primer_2023,rizk_performance_2026}.
	
	This work aims to quantify the influence of two 4-qubit Calderbank-Shor-Steane (CSS) quantum error correction codes (QECC) \cite{ranu_qkd_2022} on a vertical underwater discrete variable BB84 QKD communication, specifically determining the operational limits and performance gains in this high erasure environment.

	Photon loss (i.e., the absence of a detected photon at the receiver) in photon-based quantum communication corresponds to an erasure channel in quantum information theory, an error model in which the error location is identifiable \cite{niroula_quantum_2023}. Erasures are advantageous in the context of QECC because the receiver can often identify that information was lost and recover the original state \cite{jia_quantum_2019}. Four-qubit QECCs were introduced to correct one erasure \cite{wu_erasure_2022,grassl_codes_1997}.  
	The CSS code family  \cite{calderbank_good_1996,wang_ai-enabled_2026} has been incorporated into QKD scenarios: the 5-qubit stabilizer code \cite{jayachandran_experimental_2025}, the 7-qubit CSS code  \cite{jia_quantum_2019}, and dual-rail encoding  \cite{raouf_performance_2022}. Furthermore, for quantum secure direct communication, CSS codes have also been proposed \cite{jha_joint_2024}. 
	
	All in all, in this study, we develop and validate a comprehensive channel model for vertical quantum underwater communication. Using analytical and Monte-Carlos simulations, we assess the performance in terms of quantum bit error rate (QBER) and secure key rate (SKR) of the BB84 QKD protocol with and without QECC, depending on the signal-to-noise ratio (SNR). We subsequently establish the operational conditions under which the 4-qubit code provides significant error mitigation versus those where they introduce prohibitive overhead.
	
	\section{\label{sec:level1}The underwater chain of communication}
	
\begin{figure}
		\includegraphics{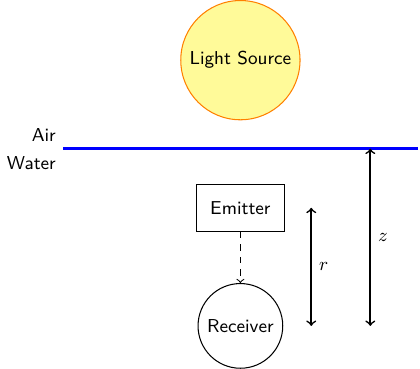} 
		\caption{Schematic of the proposed quantum communication setup for QKD. The system includes a light source in the air, a still air to water interface, a receiver underwater facing upwards located at depth $z$ below the water surface, and an emitter at a distance $r$ from the receiver also underwater}
		\label{fig:Settup}
	\end{figure}

	The chain of communication is a vertical underwater communication happening between an emitter and a receiver. The emitter is trying to communicate a secret key to the receiver. As depicted in Fig. \ref{fig:Settup}, the receiver is at range $r$ under the emitter and is located at depth $z$ from the sea surface. At the emitter the quantum state is physically encoded on photons that are then sent through the water channel to the receiver that collects the quantum state.
	Photons are generated via a none ideal single-photon source (like eDelight by Quandela) with mean photon number $\mu$ per pulse, following Poissonian statistics at repetition frequency $\xi$. The state is encoded in the polarization of these photons. The emitted photons are sent at a certain angle that follows a symmetric Gaussian distribution of a $\sigma$ angle spread. 
	
	During propagation through water, photons experience three primary degradation effects. The photons can be lost because of absorption or scattering, depending on the distance and wavelength. Furthermore, in the case where a photon was scattered because of its interactions with particles but not absorbed and arrived at the receiver,  there is a certain depolarizing effect. Lastly, a stochastic phase shift from electromagnetic interactions with the medium can arrive, bringing decoherence.
	
	For the analysis presented here, we focus exclusively on photon loss, treating it as the dominant error mechanism in the underwater regime. Incoming photons are collected during discrete time slots of duration $\Delta t$, with each slot treated as an independent quantum channel use. 
	
\begin{figure}
		\includegraphics{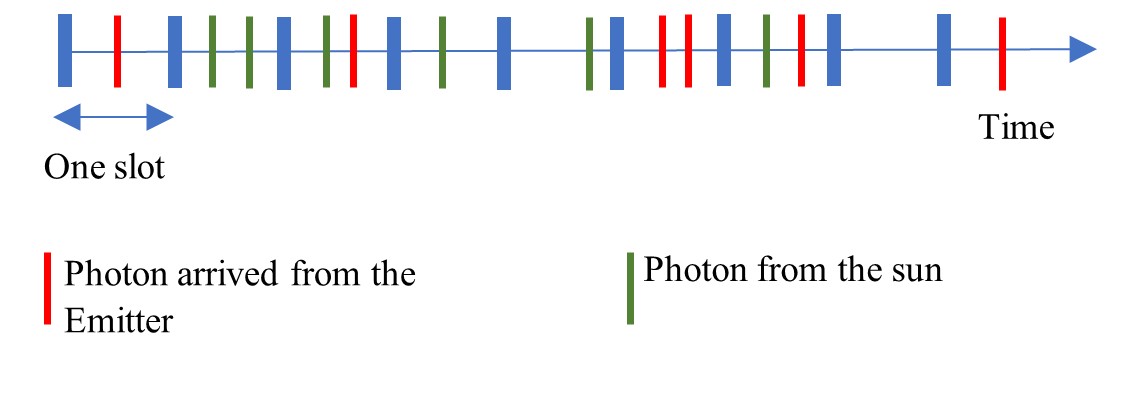} 
		\caption{ Shematic view of the time line of receiving photons over nanosecond-scale slots showing signal photons (red) and noise photons (green), each following independent Poisson distributions. The stochastic arrival patterns of both populations within individual time slots demonstrate the Poisson statistics inherent to each photon source.}
		\label{fig:distrib}
	\end{figure}

	Fig. \ref{fig:distrib} demonstrates that both sent photons and noise photons (ambient background, primarily solar radiation) arrive stochastically and independently, following Poisson distributions with expected rates in photons per slot of $\langle F \rangle$ for sent photons  and $\langle N \rangle$ for noise photons. 
	
	\begin{figure}
		\includegraphics{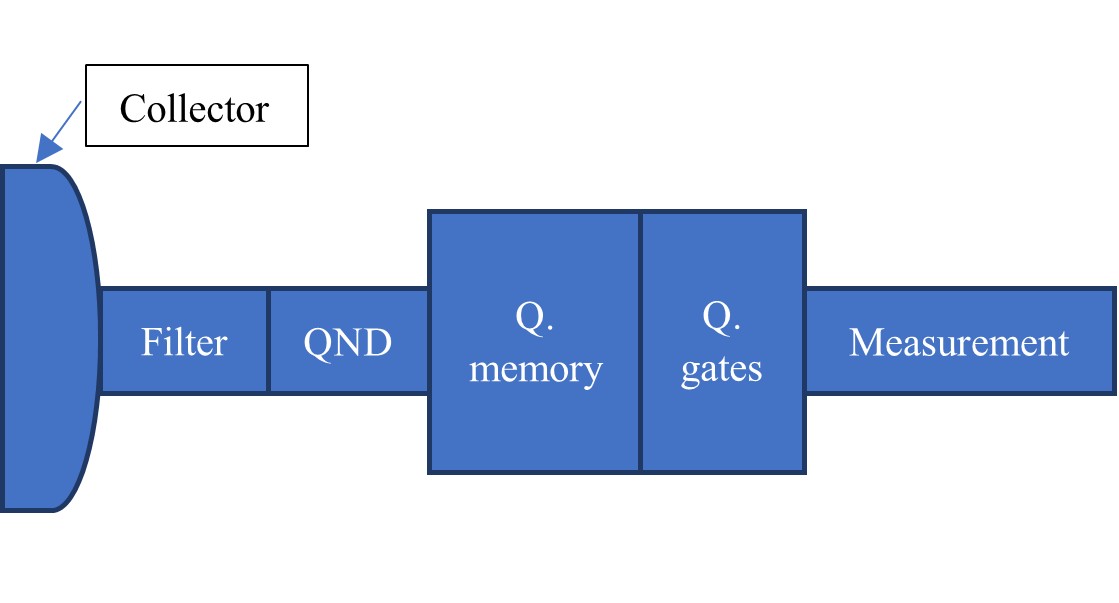} 
		\caption{Schematic of a quantum communication system at the receiver showing the signal processing pathway. Incoming photons are collected and filtered by wavelength before passing through a QND photon number detector, followed by quantum gates to apply QECC and final quantum measurement.}
		\label{fig:Settupreceiver}
	\end{figure}	
	At the receiver, as  illustrated in Fig. \ref{fig:Settupreceiver}, after collecting photons during a certain time $\Delta t$, the photons are filtered for a given wavelength $\lambda$ with a certain bandwidth $\Delta\lambda$. The wavelength of the photons is chosen as green to maximize the distance at which the photons can reach. To handle cases where multiple photons or no photons arrive, a quantum nondemolition (QND) photon number sensor is employed \cite{imoto_quantum_1985}. QND sensor allows verification of the number of photons that arrived without collapsing the polarization state. This enables the filtering of multi-photon and no-photon pulses, which is a requirement for implementing quantum error correction gates (e.g., CNOT operations) that demand exactly one photon in a photonic system. Once a single photon is confirmed, the state is stored in quantum memory. Note that, if we find a way to select one of the photons after the filter, there is no need for a QND in the case of multiple photons. Afterwards, if QECC is subsequently applied, syndrome measurements proceed; otherwise, direct measurement occurs and the QKD protocol concludes via a classical channel.

	To enable tractable analysis, we adopt the following assumptions: the water is homogenous with a refractive index of one; it is still with no scattering of noise. Moreover, the light field does not vary over the line of sight, noise photons only come from the sun and the moon and all hardware components operate with ideal fidelity. The receiver is circular and is perfectly synchronized with the emitter. We are only taking into account the geometrical loss and absorption or scattering of the photon. Finally, when there is more than one photon, we discard the slot. 
	
	\section{\label{sec:level3}Evaluation of the chain of underwater quantum communication}
	The average numbers of sent photons $\langle F \rangle$ and noise photons $\langle N \rangle$ arriving at the receiver during a detection slot of duration $\Delta t$ are determined by channel properties, emitter and receiver specifications. We derive these quantities as functions of measurable system parameters.
	
	\subsection{The average number of sent photons arriving at the collector}
	Under the assumption of weak coherent pulses with mean photon number $\mu$ and perfect emitter-receiver synchronization, the mean number of sent photons collected per slot is (see Appendix \ref{app:geometric} for derivation)
	\begin{equation} 
		\langle F \rangle = \mu \cdot p(r, D, \sigma), \label{average_sent}
	\end{equation} 
	where the probability of arrival of a sent photon at the receiver $p(r, D, \sigma)$ is the product of geometric and attenuation probabilities: 
	\begin{equation} 
		p(r, D, \sigma) = \left[1 - \exp\left(-\frac{D^2}{8r^2\sigma^2}\right)\right] \exp(-c_{\text{att}}r). 
	\end{equation} 
	The first factor represents the  geometric loss probability of arrival due to beam divergence (angle standard deviation $\sigma$) and circular receiver aperture ($D$); the second represents the probability of survival of the sent photon in a range $r$ with coefficient $c_{\text{att}}$ characteristic of the water type and wavelength.

	\subsection{The average number of noise photons arriving at the collector}
	Ambient noise photons, predominantly from downwelling solar radiation, are attenuated exponentially with depth $z$. This gives a mean number of ambient noise photons arriving during a detection slot of
	\begin{equation} 
		\langle N \rangle = \frac{E(\lambda) \cdot e^{-k_{\infty} z} \cdot \Omega_{\text{det}} \cdot S \cdot \Delta t \cdot \Delta \lambda \cdot \lambda}{h \cdot c \cdot \Omega}, \label{eq:noise_photons} 
	\end{equation} 
	where the physical parameters are defined in Table \ref{tab:table1}. 
	The derivation is detailed in Appendix \ref{app:averagenoise}. 
	
	\begin{table*}
		\caption{\label{tab:table1}Definition of the physical system parameter for the calculation of the mean number of noise photons arriving during a certain time interval with example values that will be used in the derivation of a precise example.}
		\begin{ruledtabular} 
			\begin{tabular}{lll}
				Parameter & Value and Unit & Description \\
				\colrule
				$\Omega_{\text{det}}$ & $2\pi(1-\cos(\delta/2))$ & Collector field of view \\
				$\delta$ & $180^\circ$ & Angular acceptance of the receiver \\
				$\Delta\lambda$ & $2 \times 10^{-9} m$ & Bandwidth (2 nm with OD 6 filter) \\
				$\Delta t$ & $3 \times 10^{-9} s$ & Time interval \cite{ren_ground--satellite_2017} \\
				$\lambda$ & $550 \times 10^{-9} m$ & Wavelength \\
				$D$ & $10 \times 10^{-2} m$ & Diameter of the receiver \\
				$S$ & $\pi(D/2)^2$ & Surface area of the receiver \\
				$h$ & $6.626 \times 10^{-34} J \cdot s $ & Planck constant \\
				$c$ & $3 \times 10^8 m\cdot s^{-1}$ & Speed of light \\
				$\Omega$ & $6.67 \times 10^{-5} sr$ & Solid angle of the solar disk \\
				$E(\lambda)$ & $1.85\cdot 10^{9} W m^{-2} m^{-1} $ & Spectral irradiance  \cite{mobley_light_1994} \\
				$z$ & $m$ & Depth under water \\
				$k_\infty$ &  $0.1139 m^{-1}$& Water-type dependent attenuation coefficient \cite{mobley_light_1994, miller_dynamic_2009} \\
			\end{tabular}
		\end{ruledtabular}
	\end{table*}

	Equations \eqref{average_sent} and \eqref{eq:noise_photons} specify the sent and noise photon arrival rates as functions of emitter, receiver, channel, and environmental parameters. These rates determine the signal-to-noise ratio:
	\begin{equation}
		SNR=\frac{\left\langle F\right\rangle}{\left\langle N\right\rangle}. \label{snr}
	\end{equation}

	\section{\label{sec:model_water} Quantum Channel Model and QECC}
	We represent the quantum state of a single photon in a three-dimensional Hilbert space spanned by the basis states: 
	\begin{equation*}
		|0\rangle = \begin{pmatrix} 1 \\ 0 \\ 0 \end{pmatrix}, \quad 
		|1\rangle = \begin{pmatrix} 0 \\ 1 \\ 0 \end{pmatrix}, \quad 
		|\text{vac}\rangle = \begin{pmatrix} 0 \\ 0 \\ 1 \end{pmatrix}
	\end{equation*}
	where $|0\rangle$ and $|1\rangle$ represents the two-qubit states (polarization basis), and $|\text{vac}\rangle$ denotes the vacuum (no useful quantum state) occurring when photon loss or multi-photon detection causes the state to be discarded. 
	
	To do QECC, a logical state defined in a three-dimensional Hilbert space is encoded on some physical qubit states that live in another three-dimensional Hilbert space. After defining the physical channel applied to each physical quantum state and the logical channel applied to each logical state, we will see the QECC applied. 
	
	\subsection{\label{sec:channel}The channel}
	The underwater quantum channel is modeled as a completely positive trace-preserving (CPTP) map $\varepsilon$ acting on the density matrix $\rho$ of individual physical qubits. Each physical qubit transits the channel independently. 
	
	The physical channel is the effect of the water on an individual physically encoded quantum state. The channel comprises three competing processes. 
	The first one is the transition of the state to the vacuum state $|\text{vac}\rangle$ with a probability of $g$. This occurs when no photon arrives in the slot or when multi-photon detection occurs and the state is discarded by the QND filter. 
	The second process is the successful transmission of the state with a probability of $v$. This is the case when exactly one sent photon arrives and is detected without noise, preserving the quantum state.
	The last process happens when exactly one noise photon arrives alongside zero sent photons, driving the state toward the maximally mixed state $I/2$ with a probability of $w$. The quantum channel that represents the action of the state encoded on the photon going through the underwater channel can be defined as a mixture of an erasure channel and a Pauli channel, as expressed Eq. (\ref{eq:chan1}).
	\begin{equation}
		\varepsilon\left(\rho\right)=gTr\left(\rho\right)|\text{vac}\rangle\langle\text{vac}|+ v\rho+wTr\left(\rho\right) \frac{I}{2}. \label{eq:chan1}
	\end{equation}
	The different probabilities are linked to the channel conditions. Recalling $\langle F \rangle$ denotes the mean number of signal photons and $\langle N \rangle$ the mean number of noise photons per detection slot. Assuming independent arrivals:
	\begin{subequations}
		\begin{eqnarray}
			g&=&1-(\left\langle N\right\rangle+\left\langle F\right\rangle)\mathrm{e}^{-\left\langle F\right\rangle}\mathrm{e}^{-\left\langle N\right\rangle}, \label{appa}
			\\
			v&=&\left\langle F\right\rangle\mathrm{e}^{-\left\langle F\right\rangle}\mathrm{e}^{-\left\langle N\right\rangle}, \label{appb}
			\\
			w&=&\left\langle N\right\rangle\mathrm{e}^{-\left\langle N\right\rangle}\mathrm{e}^{-\left\langle F\right\rangle} . \label{appc}
		\end{eqnarray}
	\end{subequations}
	Having $g + v + w = 1$ ensures the map is trace-preserving.

	The logical channel is defined when there is QECC. When a logical qubit is encoded via QECC across multiple physical qubits, the effective logical channel $\tilde{\varepsilon}$ acting on the logical density matrix $\rho_L$ accounts for the cumulative effect of errors on all physical qubits. Under the assumption that each physical qubit experiences the channel $\varepsilon$ independently, the logical channel becomes:
	\begin{eqnarray}
		\tilde{\varepsilon}\left(\rho_{L}\right)&=&\tilde{g}Tr\left(\rho_{L}\right)|\text{vac}\rangle_{L}\langle\text{vac}|_{L}\nonumber \\
		&+& \tilde{\alpha}\rho_{L}\nonumber \\
		&+&\frac{\tilde{w}}{4}\sum_{\substack{i,j=0 \\ j \neq i, \text{ if }i\text{ or } j=0}}^1 X_i Z_j \rho_{L} X_i^\dagger Z_j^\dagger, \label{logicalchannel}
	\end{eqnarray}
	where 
	\begin{equation*}
		X = \begin{pmatrix} 0 & 1 & 0 \\ 1 & 0 & 0 \\ 0 & 0 & 1 \end{pmatrix}, \quad
		Z = \begin{pmatrix} 1 & 0 & 0 \\ 0 & -1 & 0 \\ 0 & 0 & 1 \end{pmatrix}, \quad
		Y = \begin{pmatrix} 0 & -i & 0 \\ i & 0 & 0 \\ 0 & 0 & 1 \end{pmatrix}.
	\end{equation*}
	The parameters $\tilde{g}$, $\tilde{\alpha}$, $\tilde{w}$ have analogous interpretations: $\tilde{g}$ is the probability that the logical state is erased (lost); $\tilde{\alpha}$ is the probability of faithful transmission; and $\tilde{w}$ is the probability of a Pauli (X, Y, Z, I) error in the logical state. The CPTP condition requires $\tilde{g} + \tilde{\alpha} + (3/4)\tilde{w} = 1$. The specific values of $\tilde{g}$, $\tilde{\alpha}$, $\tilde{w}$ depend on both the physical channel parameters ($g$, $v$, $w$) and the encoding structure.
	\subsection{\label{sec:QECC}Four qubit CSS codes for erasure correction}
	In photon-based quantum communication dominated by erasure errors (photon loss or filtering), QECC optimized for erasure channels is essential \cite{niroula_quantum_2023}. We examine two variants of four-qubit CSS stabilizer codes \cite{wu_erasure_2022,grassl_codes_1997}.
	
	The standard code is a $[4, 1, 2]$ CSS stabilizer code \cite{wu_erasure_2022,grassl_codes_1997} encoding one logical qubit into four physical qubits. It can correct one erasure if its location is known and only detect if there is an error or not, as the minimal weight of a logical operator is 2 \cite{grassl_codes_1997}. The stabilizer generators are: 
	\begin{equation*}S=\ \left\langle
		\begin{matrix}g_1=XXXX\\g_2=ZIIZ\\g_3=IZZI\\
		\end{matrix}
		\right\rangle.
	\end{equation*}
	The logical basis states are:
	\begin{eqnarray*}
		|0\rangle_L &= \frac{|0000\rangle + |1111\rangle}{\sqrt{2}}, \\
		|1\rangle_L &= \frac{|1001\rangle + |0110\rangle}{\sqrt{2}}.
	\end{eqnarray*}
	The logical state is a vacuum state if there is more than one erasure on the physical qubits or if the syndrome measurement shows that there is an error.
	
	Moreover, we implemented a discard variant that gives an immediate declaration of logical vacuum $|\text{vac}\rangle_L$ if any physical qubit is erased. This variant is useful for comparison: it avoids the decoding complexity and potential error propagation of the standard code but sacrifices any benefit of single-erasure correction.
	
	\section{\label{sec:BB84}BB84 protocol to determine the impact of QECC underwater}
	We assess the impact of the 4-qubit CSS codes on BB84 QKD performance in the underwater channel \cite{bennett_quantum_2014}. Two key metrics characterize BB84 performance: the QBER and the SKR. These quantities depend directly on the physical channel parameters. 
	
	\subsection{BB84 without QECC}
	For the physical channel $\varepsilon$ given in Eq. \eqref{eq:chan1}, the QBER is defined as the ratio of erroneous sifted bits to total sifted bits. Given the channel structure and that the sifted key is the size of the whole register, the QBER is:
	\begin{eqnarray}
		QBER_{QKD}&=&\frac{w}{2\left(1-g\right)} \nonumber \\ 
		&=& \frac{\left\langle N\right\rangle}{2\left(\left\langle N\right\rangle+\left\langle F\right\rangle \right)}\nonumber \\ 
		&=& \frac{1}{2\left(1+SNR \right)},
	\end{eqnarray}
	where $SNR$ is the signal-to-noise ratio defined in Eq. \eqref{snr}.
	
	The SKR is the rate at which a secret key can be delivered. In BB84 its performance is characterized by the mutual information $I_{\text{ER}}$ between the emitter and receiver and the Holevo bound $\chi_{EA}$ quantifying an attacker's accessible information in a collective attack.
	
	The channel behaves as a binary symmetric erasure channel with an erasure probability $g$. The associated mutual information is: 
	\begin{equation} 
		I_{ER} = (1-g)\left(1-h_2\left(\frac{w}{2(1-g)}\right)\right)
	\end{equation} 
	with $h_2$ the binary entropy.
	In the absence of eavesdropping, the secret key rate is 	
	\begin{equation}
		K_{no\_sec}= \xi \cdot I_{ER} \label{SKR}.
	\end{equation}
	
	Against collective attacks, the attacker $A$ can access information at a rate quantified by the Holevo bound $\chi_{EA}$. For the BB84 protocol, assuming perfect post-processing error correction, this yields \cite{nath_tuto_2026, Devetak_2005}: 
	\begin{eqnarray}
		K_{sec} &=& I_{ER}-\chi_{EA} \nonumber\\
		&=&R(1-h_2(QBER)- h_2(QBER)),
	\end{eqnarray} 
	with $h_2$ the binary entropy function in log 2, and $R$ the raw rate of detecting a state. 
	
	In the defined physical channel $\varepsilon$, we have:
	\begin{equation}
		K_{sec} =\xi \cdot(1-g)(1-2\cdot h_2(\frac{\beta}{1-g})).\label{SKRsec}
	\end{equation}
	Security is guaranteed only when $\chi_{\text{EA}} < I_{\text{ER}}$, which imposes the well-known QBER threshold $\text{QBER} < 11\%$.
	
	Crucially, to maintain QECC validity, each successfully detected sent photon is assumed to correspond to the physical qubit encoded for that slot, establishing the necessary mapping between stochastic photon arrivals and deterministic code structure.
	
	\subsection{BB84 with 4-qubit QEC code}
	When a 4-qubit CSS code encodes the logical qubits, the BB84 protocol operates on the logical channel $\tilde{\varepsilon}$ (Eq. \eqref{logicalchannel}), which has an erasure probability $\tilde{g}$ and a Pauli error probability $\tilde{w}$. The corresponding metrics in BB84 become: 
	\begin{align}
		&QBER_{QKD} = \frac{\tilde{\beta}}{1-\tilde{g}}, \label{QBERQECC}\\
		&K_{no\_sec} = \frac{\xi}{4} \cdot(1-\tilde{g})(1-h_2(\frac{\tilde{\beta}}{1-\tilde{g}})), \label{nosecskrQECC}\\
		&K_{sec}=\frac{\xi}{4} \cdot(1-\tilde{g})(1-2\cdot h_2(\frac{\tilde{\beta}}{1-\tilde{g}})), \label{secskrQECC}
	\end{align} 
	where $\tilde{\beta} = \tilde{w}/2$ and the factor $1/4$ accounts for the 4-to-1 qubit expansion (one logical qubit encoded on four physical qubits). 
	
	The factors are computed from the BB84 ensemble via:
	\begin{eqnarray}
		\tilde{\beta} &=& \sum_{v \in \mathcal{S}} P(V=v) \sum_{v' \in \mathcal{S} \setminus \{v\}} P(V_{out}=v' \mid V=v), \label{eq:beta_logical}  \\
		\tilde{g} &=&  P(V=|0\rangle_L) \, P(V_{out}=|\text{vac}\rangle_L \mid V=|0\rangle_L) \nonumber\\
		&+& P(V=|1\rangle_L) \, P(V_{out}=|\text{vac}\rangle_L \mid V=|1\rangle_L) \nonumber\\
		&+& P(V=|+\rangle_L) \, P(V_{out}=|\text{vac}\rangle_L \mid V=|+\rangle_L) \nonumber\\
		&+& P(V=|-\rangle_L) \, P(V_{out}=|\text{vac}\rangle_L \mid V=|-\rangle_L), \label{eq:g_logical}
	\end{eqnarray}
	with $V$ the input state, $V_{out}$ the output state and $\forall v \in \mathcal{S}, P(V=v) = 1/4$ because there is an equiprobability of sending a state from 
	\begin{eqnarray*}
		\mathcal{S} &= &\{|0\rangle_L, |1\rangle_L, \\&&|+\rangle_L=(|0\rangle_L +|1\rangle_L)/\sqrt(2) , \\&& |-\rangle_L=(|0\rangle_L -|1\rangle_L)/\sqrt(2)\} .
	\end{eqnarray*}	
	The general expression in Eq. \eqref{eq:beta_logical} counts all transitions between distinct logical basis states. However, in the BB84 protocol with symmetric encoding, this simplifies to transitions between the two conjugate bases: 
	\begin{eqnarray}
		\tilde{\beta} &= & P(V=|0\rangle_L) \, P(V_{out}=|1\rangle_L \mid V=|0\rangle_L) \nonumber\\
		&+& P(V=|1\rangle_L) \, P(V_{out}=|0\rangle_L \mid V=|1\rangle_L) \nonumber\\
		&+& P(V=|+\rangle_L) \, P(V_{out}=|-\rangle_L \mid V=|+\rangle_L) \nonumber\\
		&+& P(V=|-\rangle_L) \, P(V_{out}=|+\rangle_L \mid V=|-\rangle_L). \label{eq:beta_simplified}
	\end{eqnarray}
	To compute $\tilde{g}$ and $\tilde{\beta}$ for the standard and discard codes, we enumerate all possible error patterns (erasures and Pauli errors) affecting the four physical qubits, apply the error and erasure correction procedure, and track the resulting logical state. This yields error transition probabilities for each input state, which are summed up according to Eqs. \eqref{eq:beta_logical} and \eqref{eq:g_logical}. 
	Inputting Eqs. (\ref{eq:beta_simplified}) and (\ref{eq:g_logical}) into Eqs. (\ref{QBERQECC}) (\ref{nosecskrQECC}) and (\ref{secskrQECC}) we can calculate the QBER and the SKR. Therefore, we present the SKR and QBER calculations both analytically (using Eqs. \eqref{eq:beta_simplified} and \eqref{eq:g_logical}) and via Monte-Carlo simulation (see Appendix \ref{app:simulation}).
	
	\section{\label{sec:Results} Effect of additional QECC in BB84}
	In this section, we present analytical and simulation-based results quantifying how the two 4-qubit CSS codes (standard and discard variants) affect BB84 QKD performance in underwater channels.  We first analyze the effect of additional QECC depending on the SNR and then in the case of a realistic scenario. Throughout, we identify regimes where QECC provides net benefits and scenarios where it incurs prohibitive overhead.
	
	\subsection{QBER depending on the SNR}
	\begin{figure}
		\includegraphics{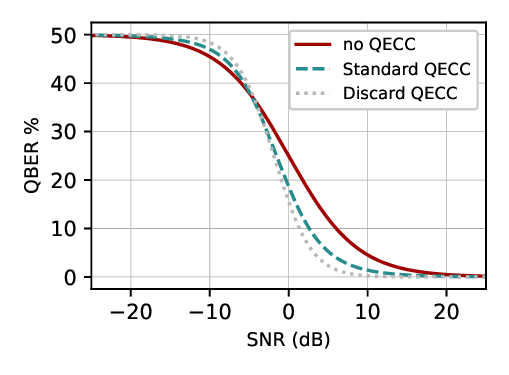} 
		\caption{QBER is plotted against SNR (dB, defined as $10 \log_{10}(\text{SNR})$) for the BB84 QKD protocol under three error-correction regimes. Red solid line: raw BB84 protocol without QECC. Teal dashed line: BB84 with standard QECC implemented on 4 physical qubits. Gray dotted line: BB84 with discard-type QECC variant, also on 4 qubits. Lower QBER indicates better error suppression and higher security margins. The standard QECC and discard QECC both reduce QBER over the raw baseline under a certain threshold, enabling key generation at SNR values where the raw protocol becomes insecure (QBER $> 11\%$). }
		\label{fig:QBERSNR}
	\end{figure}
	The QBER is the fundamental metric determining BB84 security. Fig. \ref{fig:QBERSNR} shows QBER as a function of SNR (in dB, $10 \log_{10}(\text{SNR})$, ranging from $-25$ to $25$ dB) for three configurations: raw BB84 without QECC, BB84 with standard $[4,1,2]$ CSS code, and BB84 with discard QECC variant. 
	
	Both QECC variants reduce QBER across a wide SNR range, provided that the baseline QBER (no QECC) remains below $37.4\%$ (standard code) or $36.3\%$ (discard variant). This threshold represents the crossover point beyond which the error correction overhead exceeds the error suppression benefit. Equivalently, QECC improves QBER whenever the uncorrected error rate lies below these critical values. 
	
	At the security threshold QBER $= 11\%$, the standard $[4,1,2]$ code achieves a 3 dB SNR gain, while the discard variant achieves 4.5 dB. At this critical operating point, the discard code outperforms the standard code, demonstrating superior error suppression.
	
	\subsection{SKR depending on the SNR}
	While QBER analysis confirms that QECC reduces SNR requirements, the achievable SKR reveals a more nuanced trade-off. Since multiple ($\left\langle N\right\rangle,\ p$) pairs can produce identical SNR values while differing in $1-g$ factors, SKR performance varies significantly across deployment scenarios even at fixed SNR. We examine both unsecured ($K_{\text{no-sec}}$) and secure ($K_{\text{sec}}$) key rates to quantify this dependence. 
	\begin{figure}
		\includegraphics{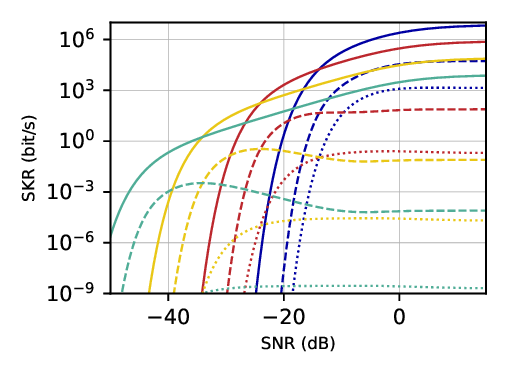} 
		\caption{SKR (bits per second) in the unsecured case $K_{no\_sec}$, where it is only linked to the capacity of the channel, is plotted against SNR (dB, $10\log_{10}(\text{SNR})$) for the BB84 QKD protocol with varying photon-arrival probabilities $p$ at the receiver. The three error-correction regimes are: solid line, no QECC; dashed line, standard QECC on 4 physical qubits; dotted line, discard-type QECC variant on 4 qubits. Photon-arrival probabilities shown are $p=1$ (ideal transmission, blue), $p=0.1$ (red), $p=0.01$ (yellow), and $p=0.001$ (teal).}
		\label{fig:SKRSNR}
	\end{figure}
	\begin{figure}
		\includegraphics{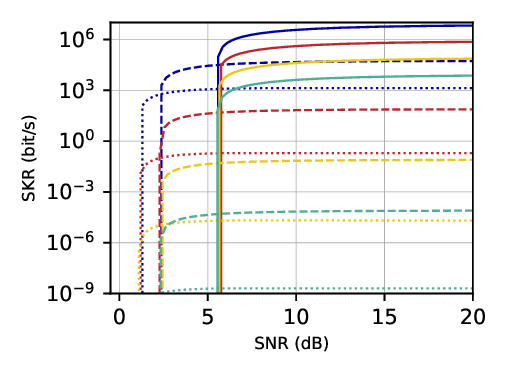} 
		\caption{SKR (bits per second), in the fully secure case $K_{sec}$, is plotted against SNR (dB, $10\log_{10}(\text{SNR})$) for the BB84 QKD protocol with varying photon-arrival probabilities $p$ at the receiver. The three error-correction regimes are: solid line, no QECC; dashed line, standard QECC on 4 physical qubits; dotted line, discard-type QECC variant on 4 qubits. Photon-arrival probabilities shown are $p=1$ (ideal transmission, blue), $p=0.1$ (red), $p=0.01$ (yellow), and $p=0.001$ (teal).}
		\label{fig:SKRSNR2h}
	\end{figure}
	
	Figure \ref{fig:SKRSNR} displays the unsecured SKR $K_{\text{no-sec}}$  versus SNR for varying photon arrival probabilities $p \in {1, 0.1, 0.01, 0.001}$ and three QECC configurations (no QECC, standard, discard). Figure \ref{fig:SKRSNR2h} shows the corresponding secure SKR $K_{\text{sec}}$. These complementary perspectives isolate the effects of QECC on information capacity (Fig. \ref{fig:SKRSNR}) versus achievable secure key generation (Fig. \ref{fig:SKRSNR2h}).
	
	Figure \ref{fig:SKRSNR} reveals non-monotonic behavior: QECC improves $K_{\text{no-sec}}$ at low SNR, particularly in the range $-40$ to $-20$ dB, where CSS codes recover key generation capacity when QBER exceeds $11\%$. This creates a characteristic enhancement "bump" near the security failure threshold. However, in this same regime, raw BB84 can tolerate lower SNR values (higher erasure rates), indicating that QECC's benefit is marginal rather than dominant.
	
	In contrast, Figure \ref{fig:SKRSNR2h} shows that below approximately $5.5$ dB SNR, secure key generation is infeasible: QBER exceeds the $11\%$ security threshold even without QECC. Standard QECC extends the feasible SNR window by $\sim 3$ dB, while discard QECC extends by $\sim 4.5$, consistent with the QBER analysis. However, this operational range extension comes at a severe cost: SKR is substantially reduced by the encoding overhead of the 4-qubit codes. 
	
	A key observation in both figures is that beyond SNR $\sim 10$ dB, both $K_{\text{no-sec}}$ and $K_{\text{sec}}$ saturate and become independent of SNR, reaching plateau values determined entirely by the photon arrival probability $p$. We now derive this high-SNR behavior analytically.
	
	Defining $s = 1 - g$ as the probability of non-erasure, we have 
	\begin{equation}
		s = 1 - g = \left(\langle F \rangle + \langle N \rangle\right) e^{-(\langle F \rangle + \langle N \rangle)}.
	\end{equation}
	
	In the high-SNR limit, noise is negligible ($\langle N \rangle \ll \langle F \rangle$), so:
	\begin{equation*}
		s = \langle F \rangle \cdot e^{-\langle F \rangle} = \mu p \cdot e^{-\mu p}.
	\end{equation*}
	For small $\mu p$ (which holds since $\mu = 0.1$), Taylor expansion yields:
	\begin{equation*}
		s \approx \mu p + \mathcal{O}((\mu p)^2).
	\end{equation*}
	At high SNR, QBER $\to 0$, so $h_2(\text{QBER}) \to 0$ in Eq. \eqref{SKR} and Eq. \eqref{SKRsec}. Thus:
	\begin{equation}
		K_{\text{no\_QECC}} \approx \xi s \approx \xi \mu p. \label{eq:K_noQECC_highSNR}
	\end{equation}
	At high SNR, noise-induced Pauli errors are negligible. Only two scenarios produce non-vacuum logical states: either all four physical qubits survive erasure (probability $s^4$), or exactly one physical qubit is erased and the CSS code corrects it (probability $4s^3(1-s)$). Therefore: 
	\begin{equation}
		\tilde{s} =  s^4 + 4s^3(1-s) = 4s^3 - 3s^4
	\end{equation}
	For $s \ll 1$:
	\begin{equation}
		\tilde{s} \approx 4s^3 \approx 4(\mu p)^3
	\end{equation}
	The corresponding secure key rate is:
	\begin{equation}
		K_{\text{standard\_QECC}} = \frac{\xi}{4}\left(4s^3 - 3s^4\right)  \approx \xi(\mu p)^3. \label{eq:K_stand_highSNR}
	\end{equation}
	For the discard variant, which discards any logical state if any physical qubit is erased:
	\begin{equation*}
		\tilde{s}_{discard} = s^4,
	\end{equation*}
	Thus:
	\begin{equation}
		K_{\text{discard\_QECC}} \approx \frac{\xi}{4} \cdot s^4. \label{eq:K_discard_highSNR}
	\end{equation}
	It is possible to verify that, numerically, in Fig. \ref{fig:SKRSNR2h} and Fig. \ref{fig:SKRSNR} at high SNR, we can find approximately the same numerical SKR as analytically using the previous approximations.

	\begin{table} 
		\caption{\label{tab:table2}%
			The ratio of SKR when reducing the probability of arrival $p$ by ${10}^{-1}$ numericaly calculated from in Fig. \ref{fig:SKRSNR2h} and Fig. \ref{fig:SKRSNR}.}
		\begin{ruledtabular}
			\begin{tabular}{ccc}
				No QECC & Standard QECC & Discard QECC \\ \colrule
				$10^{-1}$           & $10^{-3}$                 & $10^{-4}$                \\ 
			\end{tabular}
		\end{ruledtabular}
	\end{table}
	
	Table \ref{tab:table2} quantifies the SKR ratio when reducing, for the same QECC scenario, the probability of arrival by a factor of $f=0.1$: 
	\begin{equation}
		\frac{\text{SKR at } p_{reference}\cdot 0.1}{\text{SKR at } p_{reference}}
	\end{equation}
	The scaling exponents reveal the cumulative effect of encoding overhead. For raw BB84, SKR scales as $f^1$ (linear loss), reflecting direct dependence on photon arrival probability. For standard QECC, SKR scales as $f^3$, capturing the cubic dependence on single-qubit survival. For discard QECC, SKR scales as $f^4$, corresponding to the requirement that all four physical qubits must survive erasure. These empirical scalings match analytical predictions from Eq. \eqref{eq:K_noQECC_highSNR}, Eq.\eqref{eq:K_stand_highSNR} and Eq.\eqref{eq:K_discard_highSNR}. 
	
	Moreover, comparing the key rates between with and without SKR analytically, we get:
	\begin{eqnarray*}
		\Delta_{Standard QECC}^{no-QECC}SKR_{p_{target}}&=& \frac{K_{\text{standard\_QECC}}}{K_{\text{no\_QECC}}} \\ &=& \frac{4s^3 - 3s^4}{4s}= s^2\left(1 - \frac{3s}{4}\right),
		\\
		\Delta_{Discard QECC}^{no-QECC}SKR&=&\frac{K_{\text{\text{discard\_QECC}}}}{K_{\text{no\_QECC}}} = \frac{s^4}{4s}= \frac{s^3}{4}.
	\end{eqnarray*}
	
	When $s = 0.1 \cdot 0.1\approx 0.01$:
	\begin{eqnarray*}
		\Delta_{Standard QECC}^{no-QECC}SKR_{p_{target}} &\approx& 1 \cdot 10^{-4},\\
		\Delta_{Discard QECC}^{no-QECC}SKR &\approx& 2.5 \cdot 10^{-7}.
	\end{eqnarray*}
	
	\begin{table*} 
		\caption{\label{tab:table3}%
			Ratio SKR at high SNR when adding QECC for different probabilities of arrival numericaly calculated from in Fig. \ref{fig:SKRSNR2h} and Fig. \ref{fig:SKRSNR}. The ratio of standard QECC to no-QECC configurations, $\Delta_{Standard QECC}^{no-QECC}SKR$, decreases by a factor of $10^{-2}$ for every $10^{-1}$ decrease in $p$. For the discard QECC, the ratio $\Delta_{Discard QECC}^{no-QECC}SKR $ decreases by a factor of $10^{-3}$ for the same decrease in $p$.}
		\begin{ruledtabular}
			\begin{tabular}{ccccc}
				$p$ & $1$ & $0.1$ & $0.01$&$ 0.001$ \\ \colrule
				$\Delta_{Standard QECC}^{no-QECC}SKR$ & $1.14 \cdot 10^{-2}$ & $1.14 \cdot 10^{-4}$ & $1.14 \cdot 10^{-6}$ & $1.14 \cdot 10^{-8}$ \\
				$\Delta_{Discard QECC}^{no-QECC}SKR$  & $2.86 \cdot 10^{-4}$    & $2.86 \cdot 10^{-7}$    & $2.86 \cdot 10^{-10}$    & $2.86 \cdot 10^{-13}$ \\
			\end{tabular}
		\end{ruledtabular}
	\end{table*}
	
	Table \ref{tab:table3} reveals that the SKR loss in Fig. \ref{fig:SKRSNR2h} introduced by adding QECC increases as the probability of arrival decreases. For $p = 0.1$, analytical and numerical results agree within numerical precision. Discrepancies at lower $p$ arise from syndrome-based rejection (discarding states with uncorrectable errors), which is not captured by the high-SNR approximation.
	
	Fig. \ref{fig:SKRSNR} and Fig. \ref{fig:SKRSNR2h} together present a fundamental trade-off. Near the feasibility limit ($11\%$), QECC is beneficial. It lowers the required SNR (extending range or enabling operation at marginal conditions) at the acceptable cost of reduced throughput. By contrast, at higher SNR, raw BB84 dominates. QECC's encoding overhead directly degrades SKR without compensating SNR improvement, making it suboptimal. At intermediate regime ($0$–$5$ dB SNR), the trade-off is configuration-dependent. Standard QECC typically outperforms discard QECC here due to lower overhead. QECC deployment should be scenario-specific. 
	
	\subsection{Results under an experimental scenario}
	To assess practical deployment conditions, we apply the full underwater model to compute $(\langle N \rangle, p)$ as functions of operational depth $z$ and range $r$. Table \ref{tab:table1} and Table \ref{tab:table4} give the values of the different parameters that will be used, supposing the communication is, in the middle of the ocean, in a type III Jerlov water and there is a sun located at the top of the atmosphere.
	
	From these parameters, we can then derive the corresponding SKR and QBER, allowing us to evaluate system performance under realistic channel conditions. In this way, the analysis moves from general trends to a precise operational case, where the suitability of QECC can be assessed more rigorously.
	\begin{table*}
		\caption{\label{tab:table4}Definition of the physical system parameter for the calculation of the QBER and SSKR with example values that will be used in the derivation of a precise example.}
		\begin{ruledtabular}
			\begin{tabular}{lll}
				Parameter & Value and Unit & Description \\
				\colrule
				$\xi$ & $80 MHz$ & Frequency pulse of the single photon source eDelight by Quandela \\
				$\omega$ & $50 microrad$ & Laser angle divergence \\
				$\mu$ & $0.1$ & Mean photon number created by the source \\
				$D$ & $10 cm$ & Diameter of the circular collector \\
				$c_{att}$ & $0.1$ & Oceanic mean attenuation coefficient \\
			\end{tabular}
		\end{ruledtabular}
	\end{table*}
	
	\begin{figure*}
		\includegraphics{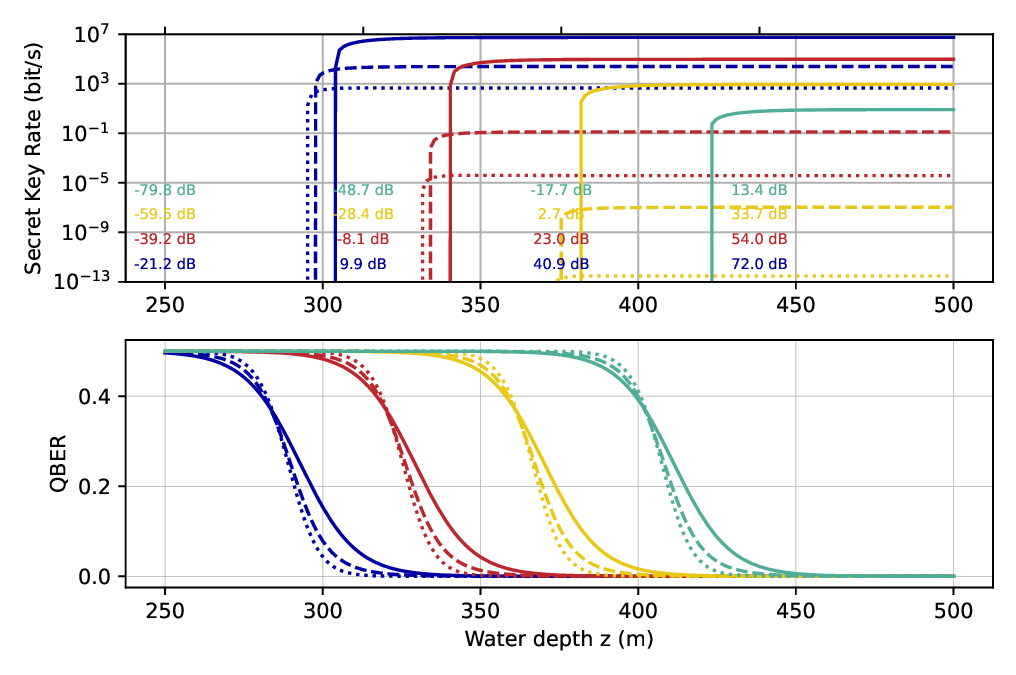}
		\caption{Fully secure against collective attacks SKR $K_{sec}$ (bits per second) and QBER plotted against communication depth $z$ (meters) for BB84 QKD at different ranges $r$ for a BB84 quantum key distribution with the sun at the top of the atmosphere with the associated SNR in the mean ocean type III. The numerical values, in the same color of the plots are the SNR values calculated at the corresponding depth for each associated range. Three QECC implementations are distinguished by line style: raw BB84 (solid line), standard QECC (dashed line), and discard QECC (dotted line). Four distinct free-space link ranges are encoded by color: $r = 3$ m (blue), $r = 43.7$ m (red), $r = 84.3$ m (yellow), and $r = 125$ m (teal). }
		\label{fig:watedepth}
	\end{figure*}
	\begin{figure*}
		\includegraphics{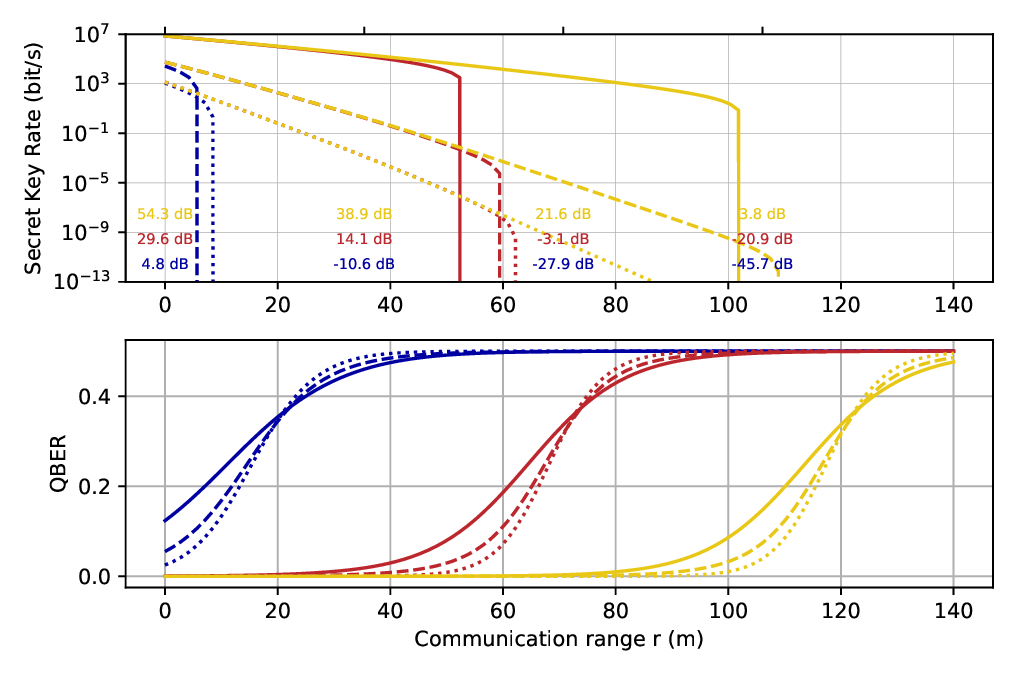}
		\caption{Fully secure against collective attacks SKR $K_{sec}$ (bits per second) and QBER are plotted against communication depth $z$ (meters) for BB84 quantum key distribution through mean ocean type III seawater with the sun at the top of the atmosphere. The numerical values, in the same color of the plots are the SNR values calculated at the corresponding depth for each associated range. Three QECC implementations are distinguished by line style: raw BB84 (solid line), standard QECC (dashed line), and discard QECC (dotted line). Three communication depths are encoded by color: $z = 300$ m (blue), $z = 350$ m (red), and $z = 400$ m (yellow).}
		\label{fig:waterange}
	\end{figure*}
	
	On the one hand, Fig. \ref{fig:watedepth} plots secure SKR $K_{sec}$ and QBER as functions of receiver depth $z$ for four fixed horizontal ranges (in meter): $r \in \{3, 43.7, 84.3, 125\}$ . On the other hand, Fig. \ref{fig:waterange} plots secure SKR and QBER vs. horizontal range $r$ for three fixed depths (in meter): $z \in \{300, 350, 400\}$ 
	As the threshold plans, below a critical depth (range-dependent) or beyond a critical range (depth dependent), QBER exceeds $11\%$ and secure key generation fails. From Fig. \ref{fig:watedepth} it is clear that the standard QECC reduces minimum depth by $\approx 6$ m, while discard QECC reduces it by $\approx 8$ m relative to raw BB84. Conversely, from Fig. \ref{fig:waterange}, the standard QECC extends the achievable range by $\approx 6.5$ m on average and discard QECC extends by $\approx 9.3$ m. These findings mean that on the terrain QECC can increase the range of communication and allow communicating at smaller depths. 
	
	However, in both these figures, it is clear that the QECC has a real consequence on the SKR and, thus, on the rate of communication. In practice, the time exchanging a key of a certain length is crucial. For example, to generate 255 bits of secure key one does not want to wait more than a maximum of a few hours; thus, the SKR should be around ${10}^{-1}$bits/s. Taking this value as an additional constraint on when communication is possible, when the sun is at the top of the atmosphere with our given experiment to have a secure QKD protocol, we can make the following decision. If the receiver is at depths of 290 meters to 320 meters; discarding QECC could be used, bringing an SKR between ${10}^2$ and ${10}^{-1}$ and ranges between 3 meters up to 20 meters. In the case of depths from 295 meters to 335 meters, standard QECC could be interesting bringing an SKR between ${10}^4$ and ${10}^{-1}$ and ranges between almost 0 meter up to 44 meters. And finally, for higher depths, No QECC is necessary to allow possible communication with high SKR.Additional QECC could even degrade too much the SKR and the effect it would have on the range is negligible compared to the order of the range possible at high depths.
	
	Adding QECC is not universally beneficial. It is most valuable in the narrow marginal SNR regime, where it enables communication otherwise impossible. In high-SNR conditions, QECC's encoding overhead dominates and degrades performance. A case-by-case assessment based on operational depth, range, and acceptable SKR threshold is essential.

	\subsection{Discussion}
	The central result of this work is robust across the explored parameter space, though several design choices merit discussion. First of all, the underwater channel model incorporates solar noise and wavelength-dependent absorption, with attenuation parameterized by $c_{\text{att}}$. Real water properties vary with salinity, temperature, pressure, and seasonal conditions; coastal and open-ocean environments exhibit different clarity profiles. Our framework accommodates this directly: practitioners can recompute the SNR-depth curve for their deployment location and reapply the QECC cost-benefit analysis. This can thereby move the SNR landscape and the QECC deployment boundary (currently 290–335 m for type III water). However, the photons sent also exhibit underwater another type of error: the stochastic phase shifts from electromagnetic interactions with the medium. In the present model, they are not resolved; incorporating them would require channel-dependent modifications to the error correction structure. This remains a near-term extension. Furthermore, the present analysis assumes unit-fidelity quantum memory, QND sensors, and logic gates. Realistic implementations incur infidelities that reduce achievable SKR and, potentially, shift the QECC cost-benefit crossover.  A finite-fidelity analysis would lower predicted key rates and potentially shift the QECC cost-benefit crossover point. Moreover, nanosecond-scale timing alignment between the transmitter and receiver is assumed. This represents a critical experimental bottleneck. Furthermore, in the scenario presented here, the vertical-upward orientation (worst case for solar noise accumulation) is fixed. Tilted or directional receivers could reduce $\langle N \rangle$ by filtering off-axis radiation, altering the SNR-noise balance. This scenario remains unexplored. In this study we also have that the state is discarded if there was more than one photon detected by the QND. As demonstrated in Appendix \ref{app:Multi}, if we could select one of the photons following a probabilistic distribution and input the state into a quantum memory, this would not change any results. Yet, on top of addressing the practical implementation of the system, in the case of QECC, further work should be done to understand how an attacker could use the structure of the QECC to eavesdrop without being detected. Despite these limitations, the central trade-off (QECC benefits at marginal SNR, overhead burden at high SNR) remains robust.

	\section{\label{sec:Conclusion}Conclusion}
	Underwater quantum key distribution via BB84 with four-qubit CSS error correction exhibits scenario-dependent benefits. For type III mean ocean water with solar background, QECC reduces the operational SNR threshold by 3 dB (standard code) or 4.5 dB (discard variant), enabling secure quantum communication at depths of 290–335 m. Beyond this marginal-feasibility window (depths $> 335 $m), raw BB84 dominates due to the SKR penalty imposed by encoding overhead and probability of arrival.
	
	These findings provide concrete engineering guidance: deploy QECC for marginal depths to lower the operational threshold; use raw BB84 for deeper water or when throughput is prioritized. The framework transparently quantifies how channel attenuation, solar noise, QECC code parameters, and implementation fidelity jointly determine SKR. It accommodates variations in water type, receiver orientation, and seasonal conditions via direct recomputation.
	
	Underwater quantum communication constitutes a critical frontier for secure global infrastructure. This work demonstrates that quantum error correction, while theoretically powerful, is not universally beneficial. Future work should be addressed coherent-attack security analysis under the assumption that adversaries know the QECC code structure; finite-fidelity gate errors; and experimental validation in realistic seawater over operationally relevant depths.
	\begin{acknowledgments}
		The authors thank Andreas ARNOLD, Thierry DEBUISSCHERT, Katia HAKEM, Ben BROWN, Nour RIZK, and Romain ALEAUME for insightful discussions on the subject of this work. 
	\end{acknowledgments}
	\appendix
	
	\section{\label{app:geometric}Derivation of the mean number of sent photons collected per slot}
	The probability of arrival 	$p(r, D, \sigma)$ of one photon depends on the probability of survival, $p_{sur}$, at distance $r$ and the probability of a photon arriving at a given location due to geometric spreading $p_{geo}$. This means, since the two actions are independent,
	\begin{equation}
		p(r, D, \sigma)=\ p_{geo}p_{sur}. \label{prob_arrival}
	\end{equation}
	
	The probability of survival $p_{sur}$ at a distance $r$ with zero scattering and zero absorption is
	\begin{equation}
		p_{sur}=\mathrm{e}^{-c_{att} r}, \label{prob_sur}
	\end{equation}
	with $c_{att}$ the total attenuation coefficient that depends on the water type and the wavelength. 
	\begin{figure*}
		\centering
		\includegraphics{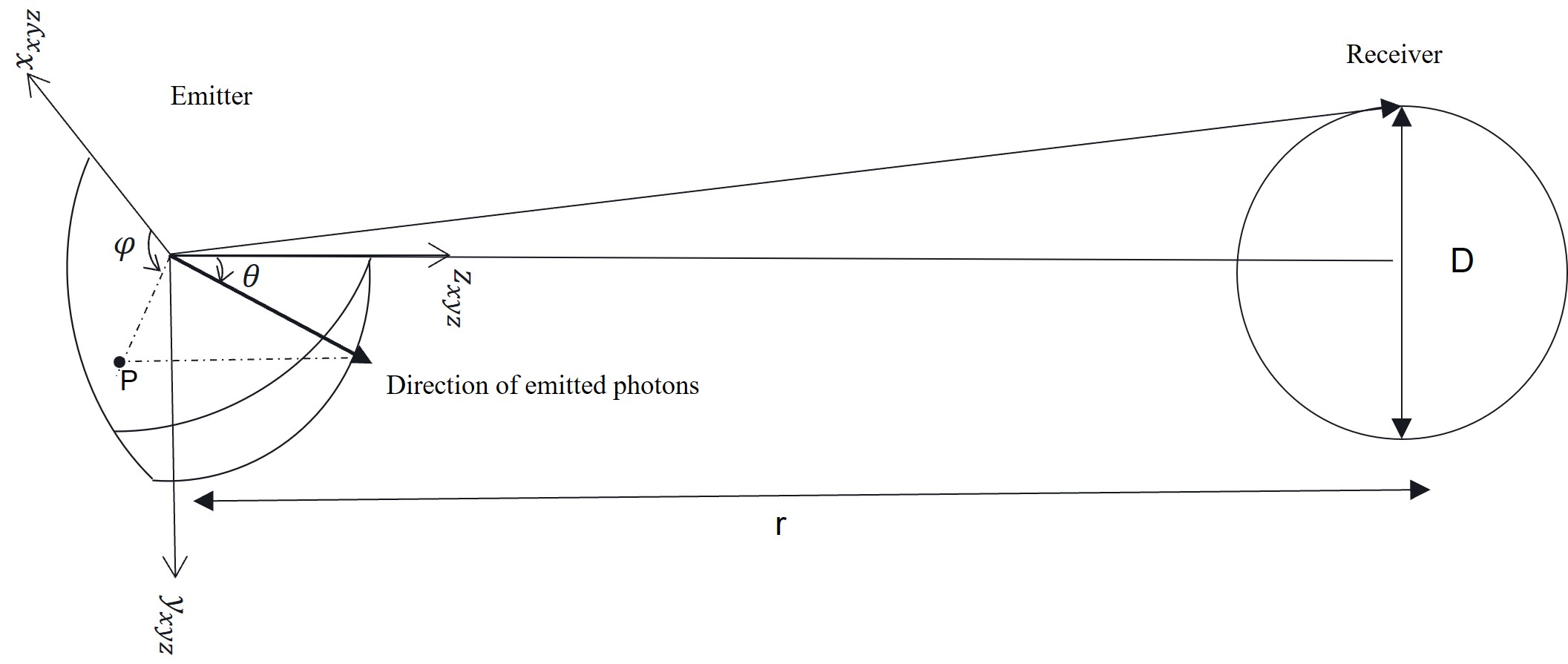}
		\caption{Shematic view of the geometrical angle scenario. An emitter (left) transmits a collimated beam with Gaussian angular divergence (half-angle standard deviation $\sigma$) toward a receiver (right) positioned at distance $r$. The receiver aperture is circular with diameter $D$. A photon emitted at angle $(\theta, \phi)$ relative to the line-of-sight direction has an angular spread described by a Gaussian distribution on the unit sphere. The projection of the direction of the sent photon  onto the $x_{xyz}y_{xyz}$ plane if noted as P. Geometric loss occurs when photons miss the receiver aperture due to the direction it was sent at; this probability depends on the ratio of receiver diameter $D$ to the beam spot size at distance $r$.}
		\label{fig:geo}
	\end{figure*}
	
	The geometrical loss in underwater is due to the angle of emission of the photon and the spreading of the transmitted beam between the emitter and the receiver \cite{elamassie_performance_2019}. Supposing that the water has a refractive index of one (air/vacuum), we can derive the probability of being detected at the distance r with a circular receiver with an aperture diameter $D$ as illustrated in Fig. \ref{fig:geo}.
	
	The probability of a photon being emitted in a direction, $p_{emission\_angle}$, defined by a polar angle $\theta$ and azimuthal angle $\phi$, as visible in Fig. \ref{fig:geo}, can be derived. Supposing that we are in a spherical coordinates based on the emitter where the $z_{xyz}$ axis is in the direction of the reseptor. $\theta$ is the angle regarding $z_{xyz}$ and $\phi$ is the angle regarding $x_{xyz}$ in the plane of $x_{xyz}y_{xyz}$. The symetry of the system regarding $z_{xyz}$ allow to define the probability density function only based on $\theta$. It is given in Eq. (\ref{emissionangle}), where $\sigma$ is the standard derivation for both angle spreads. 
	\begin{equation}
		p_{emission\_angle}\left(\theta,\ \phi\right)=\frac{1}{2\pi\sigma^2}\mathrm{e}^\frac{{-\theta}^2}{2\sigma^2} \label{emissionangle}
	\end{equation}
	Supposing that the receiver is at distance $r$ with a circular aperture of diameter $D$, the maximum accepted emitted angle is
	\begin{equation*}
		\alpha_{max}=\frac{D}{2r}. 
	\end{equation*}
	To have the probability of being at the surface of the receiver, we have
	\begin{equation}
		p_{geo}\left(D,\ r\right)=\int_{\phi=0}^{\phi=2\pi}\int_{\theta=0}^{\theta=\alpha_{max}}{\frac{1}{2\pi\sigma^2}e^{-\frac{\theta^2}{2\sigma^2}}\theta d\theta d\phi}.
	\end{equation}
	Since
	\begin{equation*}
		\int_{0}^{2\pi}d\phi=2\pi,
	\end{equation*}
	we have
	\begin{equation}
		p_{geo}\left(D,\ r\right)=\int_{\theta=0}^{\theta=\alpha_{max}}{\frac{\theta}{\sigma^2}e^{-\frac{\theta^2}{2\sigma^2}}d\theta}.\label{pgeogood}
	\end{equation}
	Changing the variables in Eq. (\ref{pgeogood}) and calculating the integral, we get geometric loss probability, 
	\begin{equation}
		p_{geo}\left(D,\ r\right)=1-\ e^{-\frac{{\alpha_{max}}^2}{2\sigma^2}} \label{prob_geo}
	\end{equation}
	with
	\begin{equation}
		\alpha_{max}=\frac{D}{2r}. 
	\end{equation}
	Inputting Eqs. (\ref{prob_geo}) and (\ref{prob_sur}) into Eq. (\ref{prob_arrival}) and then into Eq. (\ref{average_sent}), the average number of sent photons arriving is
	\begin{equation}
		\left\langle F\right\rangle = \mu\cdot(1-\mathrm{e}^{-\frac{{\alpha_{max}}^2}{2\sigma^2}})\mathrm{e}^{-cr}.
	\end{equation}
	\section{\label{app:averagenoise}Derivation of the average number of noise photons per slot}
	As the receiver is facing upwards, the average expected of the number of noise photons during a slot of time $\Delta t$ is 
	\begin{equation}
		\left\langle N \right\rangle = \frac{R(z, \lambda) \cdot \Omega_{det}\cdot S \cdot \Delta \lambda \cdot \Delta t \cdot \lambda}{h\cdot c}
	\end{equation}
	with, $R(z, \lambda)$ the spectral radiance in $W\cdot m^{-2}\cdot sr^{-1}\cdot  m^{-1}$ that depend on the depth $z$ and the wavelength $\lambda$ in $m$, $\Omega_{det}$ the collector field of view, $ S $ is the surface of the collector,  $\Delta \lambda $ is the bandwidth of the filter in $m$, $\Delta t$ is the time period of a slot in $s$, $h$ is Planck's constant, and $c$ is the speed of light in a vacuum. 
	
	The spectral radiance $R(z, \lambda)$ model applied in this analysis is based on the following physical assumptions \cite{miller_dynamic_2009} \cite{mobley_light_1994} :
	\begin{itemize}
		\item 	Background noise is dominated by solar irradiance. 
		\item 	The sun is treated as a uniform and isotropic source within its solid angle.
		\item   The spectral radiance at the water surface is dominated by the direct solar irradiance.
		
		\item 	The optical medium is homogeneous with a still air-water interface.
		\item 	Light field does not vary over a given horizontal plane 
		\item 	The detector is oriented upward, representing a worst-case scenario for solar noise.
	\end{itemize}
	While a comprehensive mathematical model for the evolution of radiance depending on the depth exists \cite{miller_dynamic_2009}, the empirical coefficients for scattering effects are not fully defined in the cited literature \cite{miller_dynamic_2009} \cite{mobley_light_1994}. To address this, we utilize a simplified model. This simplified model is valid if we either consider negligible scattering or consider a depth greater than 10 meters. In the case of the second hypothesis, the simplified model at a small depth gives an upper bound on the radiance value. It overestimates radiance at small depths. One can even calculate the distance from which beforehand the real radiance value is smaller \cite{miller_dynamic_2009}. This simplification provides an upper bound for radiance and maintains physical validity in deep-water scenarios.
	
	For given a wavelength, irradiance diffuse attenuation coefficient $k_{\inf}$ and depth of the communication $z$, the spectral radiance is modeled as:
	\begin{equation*}
		R(z, \lambda)=R(0, \lambda)\mathrm{e}^{-k_{\inf}\cdot z}
	\end{equation*}
	where the incident spectral solar radiance $R(0, \lambda)$ is defined as,
	\begin{equation}
		R(0, \lambda)=\frac{E(\lambda)}{\Omega_{\text{sun}}}. \label{incid_spect}
	\end{equation}
	In practice, the sea surface acts as a radiometric boundary that transforms
	the incident solar and sky field through refraction, Fresnel reflection,
	and the distribution of local surface slopes. A rough sea surface can therefore spread the transmitted light over a substantially broader angular	range than the refracted solar disk alone. Volume scattering in the water further modifies this angular distribution with depth. Consequently, the approximation Eq. \ref{incid_spect} is not generally universal. It corresponds to limiting angular models (diffuse Lambertian and direct-sun, respectively). To be more realistic, a physically predictive background model must instead specify, or experimentally calibrate, the angular radiance field seen by the receiver.
	
	In Eq. \eqref{incid_spect}, $E(\lambda)$ in $W\cdot m^{-2}\cdot  m^{-1}$ represents the solar spectral irradiance accounting for wavelength and atmospheric conditions, and $\Omega_{\text{sun}}$, in $sr$, denotes the solid angle of the solar disk as observed from the surface of the water [8].  
	This means that the number of average photons that arrive at the receiver during $\Delta t$ is
	\begin{equation}
		\left\langle N \right\rangle = \frac{E(\lambda) \mathrm{e}^{-k_{\inf}\cdot z} \cdot \Omega_{\text{det}}\cdot S \cdot \Delta \lambda \cdot \Delta t \cdot \lambda}{\Omega_{\text{sun}}\cdot h\cdot c} \nonumber.
	\end{equation}
	\section{\label{app:simulation}Monte-Carlo simulation of the BB84 with and without QECC}
	We implement a numerical simulation of the BB84 protocol to empirically deduce performance metrics. This simulation allows for the validation of our analytical results against varying channel conditions.
	\subsubsection{BB84 without QECC}
	Without QECC, the logical state is encoded only on one physical photon and thus each state goes through the channel defined in Eq.\eqref{eq:chan1}. Simulations were performed in Python by density matrix state evolution. For each of $n_{\text{shots}}$ trials, we:
	\begin{enumerate}
		\item Randomly select the emitter's bit $b_E \in \{0,1\}$ and basis $X$ or $Z$.
		\item Prepare the corresponding state in the 3-dimensional space and take the density matrix $\rho$ of it;
		\item Apply the channel $\epsilon(\rho)$ in Kraus notation to simulate transmission.
		\item The probability of measuring outcome $k, p\left(k\right)$, is calculated using the Born rule: $p\left(k\right)=Tr\left(P_k\varepsilon\left(\rho\right)\right)$, where $P_k$ is the projection onto outcome k. A Monte Carlo method randomly selects one outcome based on these probabilities.
		\item Outcomes are classified as either vacuum events, correct detections, or errors. The simulation accumulates counts: $n_{vac}$ (vacuum events) and $n_{error}$ (measurement mismatches).
		\item Accumulate statistics across all shots to calculate the QBER and SKR as such
	\end{enumerate}
	\begin{eqnarray}
		QBER&=& \frac{n_{error}}{n_{shots}-n_{vac}}, \nonumber\\
		K_{sec}&=&\xi\cdot\frac{n_{shots}-n_{vac}\ }{n_{shots}}\left(1-{2 \cdot h}_2\left(QBER\right)\right). \nonumber
	\end{eqnarray}
	The simulation reflects the asymptotic secure key rate under collective attacks.
	\subsubsection{BB84 with QECC}
	To evaluate the performance gains provided by QECC in underwater BB84, we implement a Monte Carlo simulation that models the complete communication pipeline in Qiskit.
	For each of $n_{shots}$  iterations, the simulation executes the following sequence:
	\begin{enumerate}
		\item 	Randomly select the emitter's bit and basis. This gives a random logical BB84 state (encoded in either the Z-basis or X-basis with 50\% probability each). Encode the state into four physical qubits defined by the QECC via quantum gates.
		\item 	Each of the four physical qubits traverses the underwater channel independently. Photon erasure events are stochastically selected following the probability distribution governed by g. Physically, this corresponds to a QND photon-number sensor identifying whether each qubit was successfully transmitted or erased upon arrival at the receiver. Note that a case where photons were never sent is possible as the physical photon was created with a mean number of created photons of $\mu=0.1$. The four physical qubits are collected sequentially and stored in quantum memory as they arrive. If more than one qubit is detected as erased, the logical state is immediately marked as a vacuum event and the trial is discarded. For qubits that successfully arrive, those corresponding to erasure events are replaced with a 0 state.
		\item 	Independent Pauli errors (I, X, Z, or XZ) are applied to each arriving qubit that are not erased with probability $w$. 
		\item 	If there was one erasure the corresponding erasure correction is performed depending on the location of the erasure.
		\item 	Syndrome measurement is performed via CNOT gates. The syndrome reveals the location and type of errors without collapsing the logical state. If the syndrome measurement yields a non-zero result, the trial is marked as a vacuum event.
		\item 	The left logical state is measured in the appropriate basis. The measured outcome is compared against the originally prepared logical state. The algorithm tallies vacuum events $n_{vac}$ and bit errors $n_{error}$, then computes the QBER, and SKR.
	\end{enumerate}

	\section{\label{app:Multi}Changing the Multi-photon scenario}
	The initial channel model developed assumes that time slots containing multiple photons (whether sent or noise) are immediately discarded. This represents a conservative baseline: quantum memory rejects any ambiguous state. However, a more optimistic scenario is physically imaginable: if the quantum memory can select a single photon from an arrival multi-photon case and retain only its polarization state, the receiver could recover capacity lost to multi-photon events.
	
	Specifically, we consider: when multiple photons (signal or noise) arrive in a single slot, the memory randomly selects one photon with uniform probability and stores its polarization. Let us see how would that effect the results.
	\subsection{Modified channel model: photon selection mechanism}
	Let $a$ denote the number of signal photons arriving in a slot, and $n$ the number of noise photons. Both follow independent Poisson distributions: $P(a) = \frac{\langle F \rangle^a}{a!} e^{-\langle F \rangle}$ and $P(n) = \frac{\langle N \rangle^n}{n!} e^{-\langle N \rangle}$.
	The new decision logic we have is, if $a + n = 0$ (no photons): vacuum state is stored. Channel outcome: erasure with probability $g$, if $a + n = 1$ (exactly one photon): that photon is stored. The channel outcome is success ($v$) if the photon is signal, or noise error ($w$) if it's noise, and if $a + n \geq 2$ (multiple photons): the memory selects one photon uniformly at random from the $a+n$ arrivals. The probability it selects a signal photon is $a/(a+n)$; probability of selecting noise is $n/(a+n)$.
	
	The effective channel probabilities become:
	\begin{subequations}
		
		\begin{eqnarray}
			g&=&\mathrm{e}^{-\left\langle F\right\rangle}\mathrm{e}^{-\left\langle N\right\rangle}, \label{apfpa}
			\\
			v&=&P(c_N) +{e}^{-\left\langle F\right\rangle}(1-\mathrm{e}^{-\left\langle N\right\rangle}), \label{apfpb}
			\\
			w&=&P(c_A) +{e}^{-\left\langle N\right\rangle}(1-\mathrm{e}^{-\left\langle F\right\rangle}). \label{apfpc}
		\end{eqnarray}
	\end{subequations}
	with, $P(c_A)$ the probability of saving in the memory a sent photon when there are more then one photon in the collector, and $P(c_N)$ the probability of saving in the memory a noise photon when there are more then one photon in the collector. These factors have the values:
	
	\begin{subequations}
		\begin{eqnarray}
			P\left(c_A\right)&=&\sum_{n,\ a\geq1}^{+\infty}{P(n)P(a)P\left(c_A\middle| n,\ a\right)}, \nonumber
			\\
			&=&\sum_{n,\ a\geq1}^{+\infty}{\frac{\left(\left\langle F\right\rangle\right)^a}{a!}e^{-\left\langle F\right\rangle}\ \frac{\left\langle N\right\rangle^n}{n!}e^{-\left\langle N\right\rangle}\ \frac{a}{n+a}\ },\nonumber
			\\
			&=&\left\langle F\right\rangle\left(\frac{1-e^{-(\left\langle F\right\rangle+\left\langle N\right\rangle)}}{\left\langle N\right\rangle+\left\langle F\right\rangle}\right)-e^{-\left\langle N\right\rangle}(1-e^{-\left\langle F\right\rangle}),\nonumber
			\\
			P\left(c_N\right)&=&\left\langle N\right\rangle\left(\frac{1-e^{-(\left\langle F\right\rangle+\left\langle N\right\rangle)}}{\left\langle N\right\rangle+\left\langle F\right\rangle}\right)-e^{-\left\langle F\right\rangle}(1-e^{-\left\langle N\right\rangle}). \nonumber
		\end{eqnarray}
	\end{subequations}
	
	This gives,
	\begin{subequations}
		\begin{eqnarray}
			g&=&\mathrm{e}^{-\left\langle F\right\rangle}\mathrm{e}^{-\left\langle N\right\rangle}, 
			\\
			v&=&\left\langle F\right\rangle\left(\frac{1-e^{-(\left\langle F\right\rangle+\left\langle N\right\rangle)}}{\left\langle N\right\rangle+\left\langle F\right\rangle}\right)
			\\
			w&=&\left\langle N\right\rangle\left(\frac{1-e^{-(\left\langle F\right\rangle+\left\langle N\right\rangle)}}{\left\langle N\right\rangle+\left\langle F\right\rangle}\right)
		\end{eqnarray}
	\end{subequations}
	
	\subsection{The effect of the new model on the BB84 with and without QECC}

	We now compare QBER and secure SKR predictions under both models for BB84 with and without QECC.
	
	\begin{figure}[H]
		\includegraphics{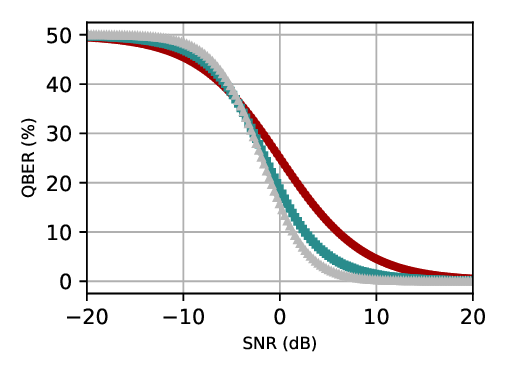} 
		\caption{QBER is plotted against SNR (dB, defined as $10 \log_{10}(\text{SNR})$) for the BB84 QKD protocol under three error-correction regimes for both discarding the multiphoton case and keeping only one photon when there are more than one. Red solid line: raw BB84 protocol without QECC. Teal dashed line: BB84 with standard QECC implemented on 4 physical qubits. Gray dotted line: BB84 with discard-type QECC variant, also on 4 qubits. If it is a line, it is the first model and if there is an additional shape (triangle, square, dot), it is the second model. Under 42\% of both models are superposed. Over 42\% of the green and red scenarios match but the blue one shows that the dashed line without shape is worse than the dashed line with squares, giving a better result with the second model. Lower QBER indicates better error suppression and higher security margins. The standard QECC and discard QECC both reduce QBER over the raw baseline under a certain threshold, enabling key generation at SNR values where the raw protocol becomes insecure (QBER $> 11\%$). }
		\label{fig:QBER_SNR2nd}
	\end{figure}
	\begin{figure*}
		\includegraphics{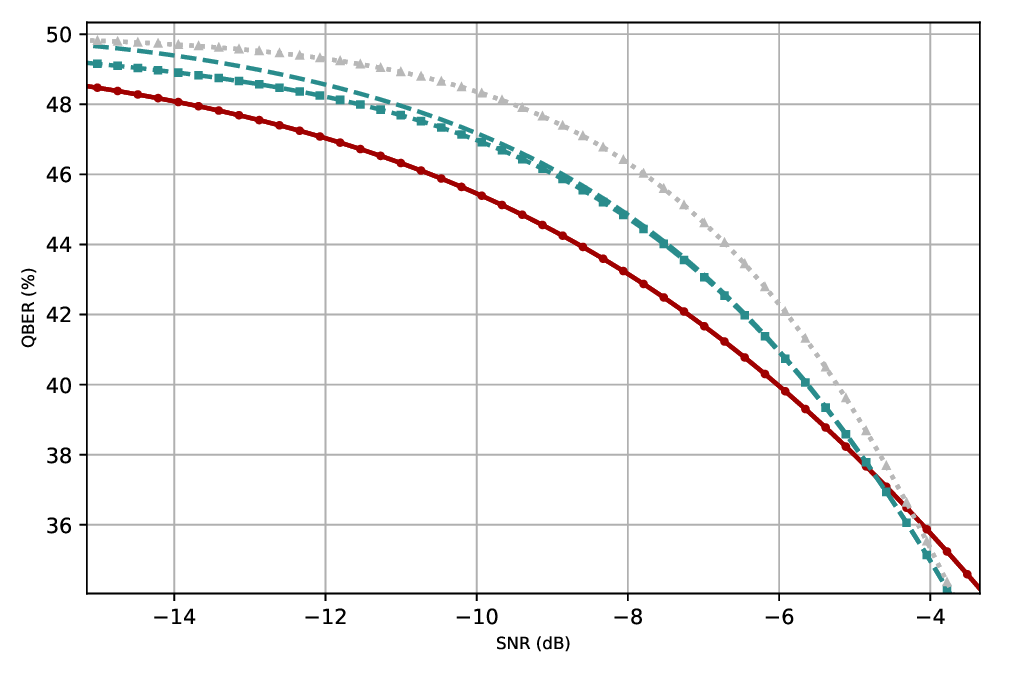} 
		\caption{QBER over 42\% is plotted against SNR (dB, defined as $10 \log_{10}(\text{SNR})$) for the BB84 QKD protocol under three error-correction regimes for both discarding the multiphoton case and keeping only one photon when there are more than one. Red solid line: raw BB84 protocol without QECC. Teal dashed line: BB84 with standard QECC implemented on 4 physical qubits. Gray dotted line: BB84 with discard-type QECC variant, also on 4 qubits. If it is a line, it is the first model and if there is an additional shape (triangle, square, dot), it is the second model. Over 42\% of the green and red scenarios match.}
		\label{fig:QBER_SNR2nd3}
	\end{figure*}
	Fig. \ref{fig:QBER_SNR2nd} shows the QBER depending on the SNR for the two different models. The first model is the one where we discard the state if there is more than one photon and the second model is the one introduced here where we keep one photon if there are more than one. It is clear that the trends of the two models are almost identical. Looking in more detail for a low SNR when the QECC does not improve communication, Fig. \ref{fig:QBER_SNR2nd3} demonstrates that over 42 \%, with a standard QECC, the second model gives the worst QBER for the same SNR. However, Fig. \ref{fig:QBER_SNR2nd} demonstrates that, below the threshold where additional QECC improves the QBER, both models give the exact QBER. 
	\begin{figure}[H]
	\includegraphics{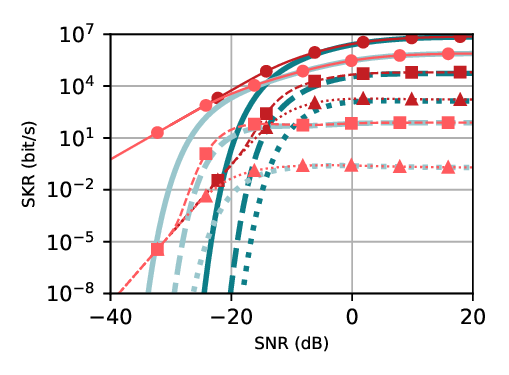} 
	\caption{SKR, in the unsecured case $K_{no\_sec}$, is plotted against SNR (dB, $10\log_{10}(\text{SNR})$) for the BB84 QKD protocol with varying photon-arrival probabilities $p$ at the receiver and for the two models. The first model is when we discard the multiphoton case and the second is when we can select one photon from the multiphoton case. The three error-correction regimes are: solid line, no QECC; dashed line, standard QECC on 4 physical qubits; dotted line, discard-type QECC variant on 4 qubits. The models are represented as teal for the first model and red for the second model. Photon-arrival probabilities shown are $p=1$ (ideal transmission,dark colors), and $p=0.1$ (light colors).}
	\label{fig:SKR_SNR_2mod}
\end{figure}

\begin{figure}[H]
		\includegraphics{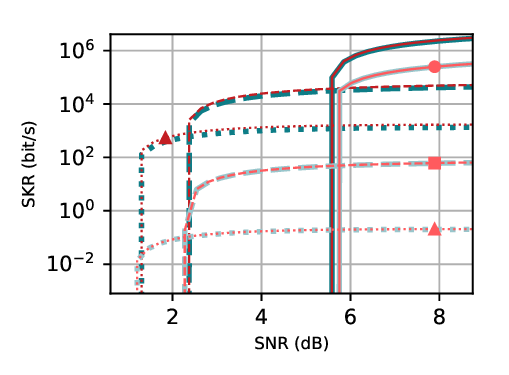} 
		\caption{SKR, in the fully secure case $K_{sec}$, is plotted against SNR (dB, $10\log_{10}(\text{SNR})$) for the BB84 QKD protocol with varying photon-arrival probabilities $p$ at the receiver and for the two models. The first model is when we discard the multiphoton case and the second is when we can select one photon from the multiphoton case. The three error-correction regimes are: solid line, no QECC; dashed line, standard QECC on 4 physical qubits; dotted line, discard-type QECC variant on 4 qubits. The models are represented as teal for the first model and red for the second model. Photon-arrival probabilities shown are $p=1$ (ideal transmission,dark colors), and $p=0.1$ (light colors). }
		\label{fig:SKR_SNR2hmodeltwo}
	\end{figure}

	Looking in terms of SKR, Fig. \ref{fig:SKR_SNR2hmodeltwo} illustrates the secure SKR depending on the SNR for the two models. The two models in the case of no QECC and with discard QECC show no differences. Yet, the case of standard QECC does show a minor improvement in SKR. This, however, is very negligible, and it is clear that both models give the same output whether you add QECC or not. Working with the unsecured SKR in Fig. \ref{fig:SKR_SNR_2mod} shows additional information. After the 11\% QBER, both models give clear different results. The second model gives a higher SKR for the same SNR. Moreover, in this second model, below $-31$ dB, whatever the probability of arrival $p$ if there is no QECC, the SKR values are the same. When there is a QECC, standard or discard case, under $-31$ dB, the SKR is equal to whatever the probability of arrival p. At high SNR the action of the second model matches the first one. 
	All in all, the second model is equivalent to the first one in the secure SKR case and for the QBER, where the QECC is interesting to add.

	\bibliography{JF_biblio.bib}
	
\end{document}